# Optimized Hierarchical Power Oscillations Control for Distributed Generation Under Unbalanced Conditions

Peng Jin [a,1], Yang Li [b,*,1], Guoqing Li [b], Zhe Chen [c], Xiaojuan Zhai [b]

[a] State Grid Customer Service Center, Tianjin 300309, P.R. China
[b] School of Electrical Engineering, Northeast Dianli University, Jilin 132012, P.R. China
[c] Institute of Energy Technology, Aalborg University, Aalborg DK-9220, Denmark

*Abstract*—Control structures have critical influences on converter-interfaced distributed generations (DG) under unbalanced conditions. Most of previous works focus on suppressing active power oscillations and ripples of DC bus voltage. In this paper, the relationship between amplitudes of the active power oscillations and the reactive power oscillations are firstly deduced and the hierarchical control of DG is proposed to reduce power oscillations. The hierarchical control consists of primary and secondary levels. Current references are generated in primary control level and the active power oscillations can be suppressed by a dual current controller. Secondary control reduces the active power and reactive power oscillations simultaneously by optimal model aiming for minimum amplitudes of oscillations. Simulation results show that the proposed secondary control with less injecting negative-sequence current than traditional control methods can effectively limit both active power and reactive power oscillations.

## I. Introduction

With the rapid development of renewable energy resources, the concept of converter-interfaced distributed generation (DG) is playing an important role in power grids [1-3]. The voltage-sourced converter (VSC) has become a key component of an energy-conversion device for DGs [4]. Due to inertialess factors of VSC, grid disturbances have adverse impact on performances of DG converters [5-6]. Most of DGs are located at the terminals of a distribution network or microgrid where unbalanced conditions exist owing to single-phase loads and asymmetrical faults [7-8]. Grid imbalance in a three-phase system leads to double working frequency power oscillations. For one thing, active power oscillations have negative effects on DC-link of converters; for another, reactive power oscillations may result in high power loss and over-current stress [9]. Consequently, various control structures have been proposed to enhance the operation performance of VSC under asymmetrical conditions in recent years.

It's known that control structures play a very important role in the DG's behavior under unbalanced conditions [10]. Notch filters are adopted to separate the components of positive- and negative-sequences from the sampling electrical quantities and the dual PI controllers in dual synchronous rotating frames (SRFs) are proposed to regulate positive- and negative-sequence components respectively [11]. The DC-link voltage ripple and active power oscillations can be suppressed by means of the dual current controllers, but reactive power oscillations are significantly amplified and system dynamics are limited by notch filters. On the contrary, to extract symmetrical components the delay signal cancellation (DSC) is adopted based on 1/4 working frequency cycle delay and combination with the original AC value [12]. Moreover, positive- and negative-sequences detection can be replaced when resonant controllers are adopted and it is demonstrated that the stationary frame resonant controllers are suitable for extracting symmetric-sequence electrical quantities. The dual SRFs can be simplified to one stationary frame and realize zero error tracking. Further, in order to promote DG performances and system dynamic tracking, controllers must be able to extract the positive- and negative-sequences components and achieve feedback control [13-14]. Therefore, all these control schemes focus on eliminating the active power oscillations and improving reference tracking, while there is seldom optimized operation strategy considering reactive power compensation and DC bus voltage oscillation simultaneously [15].

Besides control structures, it is also very important to generate current references under unbalanced conditions. Considering the constraints of DG converter, an accepted constant DC-link bus voltage should be maintained. Based on the input positive- and negative-sequence components, an unbalanced transfer matrix of input phase voltages is generally feasible [11]. However, a constant DC bus is obtained at the expense of asymmetrical currents and a sharp increase of reactive power oscillations. As discussed in [16], several schemes may improve different quality of electric energy at the point of common coupling (PCC) in terms of power oscillations and current distortions. These strategies show flexible adjustment of converters under asymmetrical voltage. However, most of them only deal with some specific distortions. It is reasonable to regulate the amplitudes of active and reactive power oscillations through an adjustable parameter of current reference [17]. Moreover, the relationship between active and reactive power oscillations is discussed, but it is only lagging in the qualitative analysis stage.

* Corresponding author. E-mail address: liyang@nedu.edu.cn (Yang Li).
[1] Peng Jin and Yang Li contributed equally to this work and should be considered co-first authors.

Previous studies mainly focus on the reduction of active power oscillations, but the analytic relationship between active and reactive power oscillations is still unclear. In addition, most work in the field of DG operation is under unbalanced grid faults, and seldom work focuses on long-term operation under a low voltage unbalance factor. However, this working condition widely exists at terminals of distribution networks or microgrids with DGs.

In this paper, a novel hierarchical control structure is proposed to suppress the power oscillations for DGs under unbalanced conditions. The contribution of this paper is two-fold: the relationship between amplitudes of the active power oscillations and the reactive power oscillations is firstly deduced and a hierarchical control structure of DGs is proposed to reduce power oscillations; and significant performance improvements from applying the proposed scheme is demonstrated in detail.

The remainder of this paper is organized as follows. First, the mechanism of power oscillations is revealed from viewpoint of the positive- and negative-sequence current injection. Then, an optimization model for suppression of power oscillations is established, and a hierarchical control structure is proposed to reduce both the active and reactive power oscillations simultaneously. Application of the proposed control scheme is demonstrated using simulation tests, and finally the conclusions are made.

## II. Mathematical Description of Converter Under Unbalanced Conditions

An unbalanced three-phase input voltage $\{E_a, E_b, E_c\}$ at PCC without a zero sequence can be represented as the sum of positive- and negative-sequences, such that

$$E_{\alpha\beta} = e^{j\omega t}E_{dq+}^+ + e^{-j\omega t}E_{dq-}^- \tag{1}$$

where $E_{dq+}^+ = E_{d+}^+ + jE_{q+}^+$, $E_{dq-}^- = E_{d-}^- + jE_{q-}^-$; +, - respectively denote the positive and negative-sequence component; $\omega$ is the angular frequency.

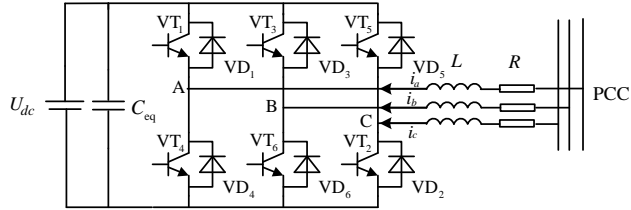

**Fig. 1.** Equivalent circuit of converter for DG

The converter model of DG is described in Fig.1 and can be decomposed as two separate parts

$$\begin{cases} E_{dq+}^+ = V_{dq+}^+ + L\dfrac{dI_{dq+}^+}{dt} + j\omega L I_{dq+}^+ + RI_{dq+}^+ \\ E_{dq-}^- = V_{dq-}^- + L\dfrac{dI_{dq-}^-}{dt} - j\omega L I_{dq-}^- + RI_{dq-}^- \end{cases} \tag{2}$$

where input current $\{I_a, I_b, I_c\}$ and output voltage of inverter $\{V_a, V_b, V_c\}$ are also expressed in forms of positive- and negative-sequence components like three-phase input voltage. With the unbalanced input voltage, apparent power is given by

$$S = (e^{j\omega t}E_{dq+}^+ + e^{-j\omega t}E_{dq-}^-)(e^{j\omega t}I_{dq+}^+ + e^{-j\omega t}I_{dq-}^-)^* \tag{3}$$

An unbalanced three-phase input voltage $\{E_a, E_b, E_c\}$ at the PCC causes double working frequency power oscillations, and instantaneous power of a DG can be expressed as

$$P(t) = P_0 + P_{c2}\cos(2\omega t) + P_{s2}\sin(2\omega t) \tag{4}$$
$$Q(t) = Q_0 + Q_{c2}\cos(2\omega t) + Q_{s2}\sin(2\omega t) \tag{5}$$

where $P_{c2}, P_{s2}, Q_{c2}, Q_{s2}$ caused by unbalanced input voltage appear as the double working frequency oscillations, and they can be expressed as

$$\begin{cases} P_0 = E_{d+}^+ I_{d+}^+ + E_{q+}^+ I_{q+}^+ + E_{d-}^- I_{d-}^- + E_{q-}^- I_{q-}^- \\ P_{c2} = E_{d+}^+ I_{d-}^- + E_{q+}^+ I_{q-}^- + E_{d-}^- I_{d+}^+ + E_{q-}^- I_{q+}^+ \\ P_{s2} = E_{d+}^+ I_{q-}^- - E_{q+}^+ I_{d-}^- - E_{d-}^- I_{q+}^+ + E_{q-}^- I_{d+}^+ \end{cases} \tag{6}$$

$$\begin{cases} Q_0 = E_{q+}^+ I_{d+}^+ - E_{d+}^+ I_{q+}^+ + E_{q-}^- I_{d-}^- - E_{d-}^- I_{q-}^- \\ Q_{c2} = E_{q+}^+ I_{d-}^- - E_{d+}^+ I_{q-}^- + E_{q-}^- I_{d+}^+ - E_{d-}^- I_{q+}^+ \\ Q_{s2} = E_{d+}^+ I_{d-}^- + E_{q+}^+ I_{q-}^- - E_{d-}^- I_{d+}^+ - E_{q-}^- I_{q+}^+ \end{cases} \quad (7)$$

Instantaneous active and reactive power in (4) and (5) can be rewritten as

$$P(t) = P_0 + P_v \sin(2\omega t + \tan^{-1}\left(\frac{P_{c2}}{P_{s2}}\right)) \quad (8)$$

$$Q(t) = Q_0 + Q_v \sin(2\omega t + \tan^{-1}\left(\frac{Q_{c2}}{Q_{s2}}\right)) \quad (9)$$

where $P_v = \sqrt{P_{s2}^2 + P_{c2}^2}$, $Q_v = \sqrt{Q_{s2}^2 + Q_{c2}^2}$. From the equations (8) and (9), the instantaneous power can be divided into AC and DC components. When a microgrid contains single-phase loads and sources, active and reactive power oscillations exist simultaneously because of unbalance voltage in PCC. Consequently, there is an urgent need to reveal the couple relationship between them.

### III. Mechanism of Power oscillation Under Unbalanced Conditions

Based on (6) and (7), the relationship between active and reactive power oscillations can be derived as followed:

$$\begin{aligned} &P_v^2 + Q_v^2 \\ &= P_{s2}^2 + P_{c2}^2 + Q_{s2}^2 + Q_{c2}^2 \\ &= (E_{d+}^+ I_{q-}^- - E_{q+}^+ I_{d-}^- - E_{d-}^- I_{q+}^+ + E_{q-}^- I_{d+}^+)^2 + (E_{d+}^+ I_{d-}^- + E_{q+}^+ I_{q-}^- + E_{d-}^- I_{d+}^+ + E_{q-}^- I_{q+}^+)^2 \\ &+ (E_{d+}^+ I_{d-}^- + E_{q+}^+ I_{q-}^- - E_{d-}^- I_{d+}^+ - E_{q-}^- I_{q+}^+)^2 + (E_{q+}^+ I_{d-}^- - E_{d+}^+ I_{q-}^- + E_{q-}^- I_{d+}^+ - E_{d-}^- I_{q+}^+)^2 \end{aligned} \quad (10)$$

Given $A = E_{d+}^+ I_{d-}^- + E_{q+}^+ I_{q-}^-$, $B = E_{d-}^- I_{d+}^+ + E_{q-}^- I_{q+}^+$, $C = E_{d+}^+ I_{q-}^- - E_{q+}^+ I_{d-}^-$, $D = E_{q-}^- I_{d+}^+ - E_{d-}^- I_{q+}^+$, (10) can be simply denoted as follows:

$$\begin{aligned} &P_v^2 + Q_v^2 \\ &= (C+D)^2 + (A+B)^2 + (A-B)^2 + (-C+D)^2 \\ &= 2(A^2 + B^2 + C^2 + D^2) \\ &= 2\{(E_{d+}^+ I_{d-}^-)^2 + 2E_{d+}^+ I_{d-}^- E_{q+}^+ I_{q-}^- + (E_{q+}^+ I_{q-}^-)^2 + (E_{d-}^- I_{d+}^+)^2 + 2E_{d-}^- I_{d+}^+ E_{q-}^- I_{q+}^+ + (E_{q-}^- I_{q+}^+)^2 + (E_{d+}^+ I_{q-}^-)^2 \\ &\quad - 2E_{d+}^+ I_{q-}^- E_{q+}^+ I_{d-}^- + (E_{q+}^+ I_{d-}^-)^2 + (E_{q-}^- I_{d+}^+)^2 - 2E_{q-}^- I_{d+}^+ E_{d-}^- I_{q+}^+ + (E_{d-}^- I_{q+}^+)^2\} \\ &= 2\{(E_{d+}^+ I_{d-}^-)^2 + (E_{q+}^+ I_{q-}^-)^2 + (E_{d-}^- I_{d+}^+)^2 + (E_{q-}^- I_{q+}^+)^2 + (E_{d+}^+ I_{q-}^-)^2 + (E_{q+}^+ I_{d-}^-)^2 + (E_{q-}^- I_{d+}^+)^2 + (E_{d-}^- I_{q+}^+)^2\} \end{aligned} \quad (11)$$

where $P_v = \sqrt{P_{s2}^2 + P_{c2}^2}$, $Q_v = \sqrt{Q_{s2}^2 + Q_{c2}^2}$, $E^+ = \sqrt{E_{d+}^{+\,2} + E_{q+}^{+\,2}}$, $E^- = \sqrt{E_{d-}^{-\,2} + E_{q-}^{-\,2}}$, $I^+ = \sqrt{I_{d+}^{+\,2} + I_{q+}^{+\,2}}$, $I^- = \sqrt{I_{d-}^{-\,2} + I_{q-}^{-\,2}}$. $P_v$ and $Q_v$ are the amplitudes of active and reactive power oscillations, respectively. Equation (11) can be simplified as

$$P_v^2 + Q_v^2 = 2(E^- I^+)^2 + 2(E^+ I^-)^2 \quad (12)$$

In (12), power oscillations are divided into two parts: positive-sequence current and negative-sequence voltage constitute one part, and negative-sequence current and positive-sequence voltage constitute another.

In the normal state, the amplitude of negative-sequence electric quantity is much less than that of positive-sequence electric quantity. Consequently,

$$E_{d+}^+ I_{d+}^+ + E_{q+}^+ I_{q+}^+ \square \ E_{d-}^- I_{d-}^- + E_{q-}^- I_{q-}^- \quad (13)$$

$$E_{q+}^+ I_{d+}^+ - E_{d+}^+ I_{q+}^+ \square \ E_{q-}^- I_{d-}^- - E_{d-}^- I_{q-}^- \quad (14)$$

$P_0$ in (6) and $Q_0$ in (7) can be approximated as

$$P_0 \approx E_{d+}^+ I_{d+}^+ + E_{q+}^+ I_{q+}^+ \quad (15)$$

$$Q_0 \approx E_{q+}^+ I_{d+}^+ - E_{d+}^+ I_{q+}^+ \quad (16)$$

The positive-sequence currents can be written as

$$\begin{bmatrix} I_{d+}^+ \\ I_{q+}^+ \end{bmatrix} \approx \begin{bmatrix} E_{d+}^+ & E_{q+}^+ \\ E_{q+}^+ & -E_{d+}^+ \end{bmatrix}^{-1} \begin{bmatrix} P_0 \\ Q_0 \end{bmatrix} \qquad (17)$$

Equation (17) indicates that the positive-sequence currents can be approximately regarded as constants if $P_0$ and $Q_0$ are determined. Accordingly, $2(E^-I^+)^2$ in (12) can also be regarded as a constant and $P_v^2 + Q_v^2$ reaches the minimum when $I_{d-}^-$ and $I_{q-}^-$ are set to zero. Therefore, $2(E^+I^-)^2$ in (12) will constitute additional increment of the power oscillations if the negative-sequence currents are injected.

Power quality of a DG is not only affected by the oscillation amplitudes but also by $P_0$ and $Q_0$. Thus, two indexes to measure power oscillations are defined as follows:

$$p^* = \frac{P_v}{P_0} = \frac{\sqrt{P_{s2}^2 + P_{c2}^2}}{P_0} \qquad (18)$$

$$q^* = \frac{Q_v}{Q_0} = \frac{\sqrt{Q_{s2}^2 + Q_{c2}^2}}{Q_0} \qquad (19)$$

The active and reactive power oscillation rates $p^*$ and $q^*$ are the ratio of the corresponding power oscillation amplitude and the average power. Therefore, a control strategy suppressing both amplitudes and rates of power oscillations is needed.

## IV. HIERARCHICAL CONTROL SCHEME OF DG CONVERTER

### A. Primary Control of DG

The hierarchical control consists of two levels. The difference between primary control and secondary control lies in current references of the controller. The active power oscillations can be suppressed in the primary level by matrix as follows [11]:

$$\begin{bmatrix} I_{d0+}^+ \\ I_{q0+}^+ \\ I_{d0-}^- \\ I_{q0-}^- \end{bmatrix} = \begin{bmatrix} E_{d+}^+ & E_{q+}^+ & E_{d-}^- & E_{q-}^- \\ E_{q+}^+ & -E_{d+}^+ & E_{q-}^- & -E_{d-}^- \\ E_{q-}^- & -E_{d-}^- & -E_{q+}^+ & E_{d+}^+ \\ E_{d-}^- & E_{q-}^- & E_{d+}^+ & E_{q+}^+ \end{bmatrix}^{-1} \begin{bmatrix} P_0 \\ Q_0 \\ P_{s2} \\ P_{c2} \end{bmatrix} \qquad (20)$$

where $P_{s2} = P_{c2} = 0$. The active power oscillations can be eliminated in ideal state. However, the negative-sequence currents are injected and reactive power oscillations cannot be reduced. Once active power oscillations are suppressed in primary control, constant DC-link voltage is kept for equivalent capacitance of converter. But only four power coefficients in (6) and (7) can be controlled and negative-sequence currents are injected when the DG is in primary control. Equation (12) indicates that negative-sequence currents could lead to additional power oscillations term $2(E^+I^-)^2$ and reactive power oscillations are enlarged.

### B. Secondary Control of DG

Once primary control achieves stable and reactive oscillations exceed a certain range, secondary control is activated. Secondary control corrects power quality by resetting current references, so both active and reactive power oscillations can be limited simultaneously. The optimal current references of secondary control can be obtained as follows:

$$\begin{cases} \min \quad F = \sqrt{P_{s2}^2 + P_{c2}^2} + \sqrt{Q_{s2}^2 + Q_{c2}^2} \\ s.t. \qquad P_0 = P_{ref} \\ \qquad\qquad Q_0 = Q_{ref} \\ \qquad \dfrac{P_{s2}^2 + P_{c2}^2}{P_{ref}^2} - \dfrac{Q_{s2}^2 + Q_{c2}^2}{Q_{ref}^2} = 0 \end{cases} \qquad (21)$$

where $P_{ref}$ and $Q_{ref}$ are power references of the DG. The aim of (21) is to reduce amplitudes of active and reactive power oscillations simultaneously, and the optimized power oscillations can be proportional to power references by the constraints. Power coefficients in (21) are the functions of $[I_{d+}^+, I_{q+}^+, I_{d-}^-, I_{q-}^-]$, and the optimized current references can be obtained through optimization.

The optimization that contains equality constraints can be transformed into unconstrained problem by Lagrange multiplier method as follows:

$$G = F + \lambda_1(P_0 - P_{ref}) + \lambda_2(Q_0 - Q_{ref}) + \lambda_3(\frac{P_{s2}^2 + P_{c2}^2}{P_{ref}^2} - \frac{Q_{s2}^2 + Q_{c2}^2}{Q_{ref}^2}) \tag{22}$$

where $\lambda_1$, $\lambda_2$ and $\lambda_3$ are Lagrange multipliers, respectively. In order to solve unconstrained optimization, partial derivatives of the objective function $G$ is derived and optimization is equivalent to nonlinear equations as follows:

$$\begin{cases} \frac{\partial F}{\partial I_{d+}^+} + \lambda_1 \frac{\partial P_0}{\partial I_{d+}^+} + \lambda_2 \frac{\partial Q_0}{\partial I_{d+}^+} + \frac{\lambda_3}{P_{ref}^2} \frac{\partial(P_{s2}^2 + P_{c2}^2)}{\partial I_{d+}^+} - \frac{\lambda_3}{Q_{ref}^2} \frac{\partial(Q_{s2}^2 + Q_{c2}^2)}{\partial I_{d+}^+} = 0 \\ \frac{\partial F}{\partial I_{q+}^+} + \lambda_1 \frac{\partial P_0}{\partial I_{q+}^+} + \lambda_2 \frac{\partial Q_0}{\partial I_{q+}^+} + \frac{\lambda_3}{P_{ref}^2} \frac{\partial(P_{s2}^2 + P_{c2}^2)}{\partial I_{q+}^+} - \frac{\lambda_3}{Q_{ref}^2} \frac{\partial(Q_{s2}^2 + Q_{c2}^2)}{\partial I_{q+}^+} = 0 \\ \frac{\partial F}{\partial I_{d-}^-} + \lambda_1 \frac{\partial P_0}{\partial I_{d-}^-} + \lambda_2 \frac{\partial Q_0}{\partial I_{d-}^-} + \frac{\lambda_3}{P_{ref}^2} \frac{\partial(P_{s2}^2 + P_{c2}^2)}{\partial I_{d-}^-} - \frac{\lambda_3}{Q_{ref}^2} \frac{\partial(Q_{s2}^2 + Q_{c2}^2)}{\partial I_{d-}^-} = 0 \\ \frac{\partial F}{\partial I_{q-}^-} + \lambda_1 \frac{\partial P_0}{\partial I_{q-}^-} + \lambda_2 \frac{\partial Q_0}{\partial I_{q-}^-} + \frac{\lambda_3}{P_{ref}^2} \frac{\partial(P_{s2}^2 + P_{c2}^2)}{\partial I_{q-}^-} - \frac{\lambda_3}{Q_{ref}^2} \frac{\partial(Q_{s2}^2 + Q_{c2}^2)}{\partial I_{q-}^-} = 0 \\ P_0 - P_{ref} = 0 \\ Q_0 - Q_{ref} = 0 \\ \frac{P_{s2}^2 + P_{c2}^2}{P_{ref}^2} - \frac{Q_{s2}^2 + Q_{c2}^2}{Q_{ref}^2} = 0 \end{cases} \tag{23}$$

*C. Quasi Newton-Trust Region (QNTR) Method*

The optimal references of positive and negative-sequence currents can be obtained by solving nonlinear equations in (23) through iterative approaches. Considering that Newton's method is very sensitive to the value of initial approximation, it may obtain an infeasible solution even with a very good initial guess. So the QNTR method is employed for doing so here. Further, the QNTR method can be summarized as follows:

The set of nonlinear equations (23) is to be regarded as an optimization as follows:

$$\min \begin{cases} F_1(X) = f_1(X) - A_1 = 0 \\ F_2(X) = f_2(X) - A_2 = 0 \\ \vdots \\ F_i(X) = f_i(X) - A_i = 0 \\ \vdots \\ F_n(X) = f_n(X) - A_n = 0 \end{cases} \tag{24}$$

subject to $X \in R^n$

Equation (24) can be simplified to a quadratic model formed by the Taylor series of the function $F_i$, as follows:

$$\min \begin{cases} S_i(\Delta_k) = F_i(X_k) + \Delta_k^T \nabla F_i(X_k) + \frac{1}{2} \Delta_k^T \nabla^2 F_i(X_k) \Delta_k \\ s.t. \; \|\Delta_k\|_2 < r_k \end{cases} \tag{25}$$

The Taylor series of the original function can be an suitable representation around the current iterate $X_k$, and the trust region step $\Delta_k$ can minimize $S_i(\Delta_k)$ within radius $r_k$.

The QNTR method starts from an initial guess $X_0 = [I_{d0+}^+, I_{q0+}^+, I_{d0-}^-, I_{q0-}^-]$ to the optimal $X^*$. For the *k*-th iteration, a step $\Delta_k$ is calculated to make $F_i(X_k + \Delta_k)$ close to $F_i(X^*)$. If this attempt is reached, the solution is updated from $(X_k)$ to $(X_k + \Delta_k)$.

Because the Hessian matrix is very complex and it must be recalculated in each iteration, second-order Taylor series $\nabla^2 F_i(X_k)$ is difficult to obtain. In order to maintain convergence of the QNTR method and reduce the computational complexity of Hessian matrix, an approximate matrix is constructed.

The new subproblem of QNTR method represented as

$$\min \begin{cases} Q_i(\Delta_k) = F_i(X_k) + \Delta_k^T \nabla F_i(X_k) + \frac{1}{2}\Delta_k^T B_k \Delta_k \\ s.t. \ \|\Delta_k\|_2 < \gamma_k \end{cases} \quad (26)$$

where $Q_i(\Delta_k)$ is the objective function of new trust-region problem; $B_k$ is the approximate matrix and a positive definite matrix.

The radius of trust region $r_k$ is calculated in each iteration on the basis of the solution in the last iteration. The updation is based on $\sigma_k$ as follows:

$$\sigma_k = \frac{F_i(X_k) - F_i(X_k + \Delta_k)}{Q_i(0) - Q_i(\Delta_k)} \quad (27)$$

In the $k$ iteration, $B_k$ can be calculated as

$$B_k = \begin{cases} B_k - \frac{B_k \Delta_k \Delta_k^T B_k}{\Delta_k^T B_k \Delta_k} + \frac{y_k y_k^T}{y_k^T \Delta_k}, \sigma_k \geq 0.01 \\ B_k, \ \sigma_k < 0.01 \end{cases} \quad (28)$$

where $y_k = \nabla F(x_{k+1}) - \nabla F(x_k)$, $B_k$ is a positive definite matrix, and $B_0$ is an identity matrix.

If $\sigma_k$ is less than 0.01, this means that $Q_i$ does not have the tendency toward $F_i$ at $X_k$. In this case, this iteration is inappropriate; its results are not updated (i.e., $X_{k+1} = X_k$), and the radius of trust region is reduced to half in the next iteration $(r_{k+1} = r_k/2)$. If the solution of quadratic model comes towards the original problem, there are two different cases. If $\sigma_k \in [0.01, 0.5]$, the results are updated with the step $(X_{k+1} = X_k + \Delta_k)$, but the radius of trust region is not changed $(r_{k+1} = r_k)$. On the contrary, if $\sigma_k \geq 0.5$, the results are updated $(X_{k+1} = X_k + \Delta_k)$ and the trust region is enlarged $(r_{k+1} = \min(2 \times r_k, r_{max}))$ within the upper limit of $r_k$ [18].

Therefore, the Quasi Newton-trust region method can be described as follows:

Step#1) Given $X_0, B_0, r_{max} > 0, \varepsilon = 10^{-3}, r_{k0} \in [0, r_{max}], k = 0$.

Step#2) If $\|\nabla F_i(X_k)\|_2 < \varepsilon$, then stop;

Step#3) Solve (26) using Dogleg method[19] to get $\Delta_k$, and compute $\sigma_k$ by (27).

Step#4) Calculate approximate matrix $B_{k+1}$, and $r_{k+1}$ is revised by value range of $\sigma_k$.

Step#5) update $k$, go to step#2.

*D. Start Flow of Hierarchical Power Oscillation Control*

During the operation, the constraint that ensures equality of active and reactive power oscillation rates need not be considered when reactive power references of DG are assigned at zero.

When three-phase voltage of PCC is unbalance, primary control adopting current references in (20) starts to reduce power oscillations at the beginning. Once primary control achieves stable and reactive ripple exceeds a certain range, secondary control starts. The flowchart of secondary control is illustrated in Fig. 2.

As illustrated in Fig. 2, the process determines the state of primary control firstly. If the difference between the current reference and the actual measured value is less than 0.04A, the primary control becomes stable. If primary control is in steady state, amplitude of reactive oscillations and reactive oscillation rate are calculated. Once amplitude of reactive power oscillations is greater than 500Var and the power oscillation ratio is not less than 10%, secondary control that reduces active and reactive power oscillations simultaneously starts after time delay.

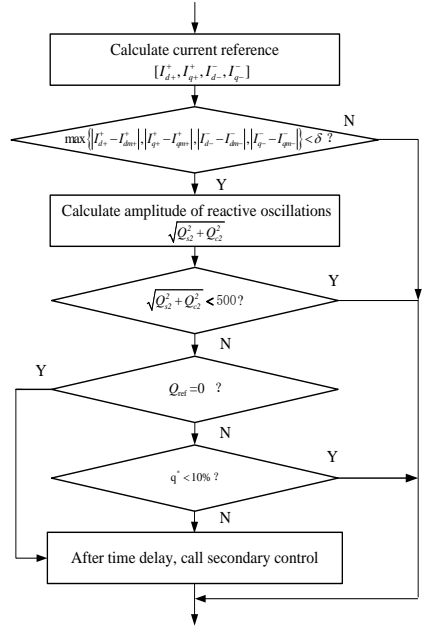

**Fig. 2.** Flowchart of secondary control

## V. CONTROL STRUCTURE OF DG

### A. Dual Current PI Controller

In order to control a voltage source converter, current references which are calculated by primary control and secondary control should be translated into voltage references as follows:

$$\begin{cases} V_{d+}^* = E_{d+}^+ - (K_p + \dfrac{K_i}{s})(i_{d+}^+ - i_{dm+}^+) + \omega L i_{q+}^+ \\ V_{q+}^* = E_{q+}^+ - (K_p + \dfrac{K_i}{s})(i_{q+}^+ - i_{qm+}^+) - \omega L i_{d+}^+ \\ V_{d-}^* = E_{d-}^- - (K_p + \dfrac{K_i}{s})(i_{d-}^- - i_{dm-}^-) - \omega L i_{q-}^- \\ V_{q-}^* = E_{q-}^- - (K_p + \dfrac{K_i}{s})(i_{q-}^- - i_{qm-}^-) + \omega L i_{d-}^- \end{cases} \quad (29)$$

where $K_p$ is the proportional gain, $K_i$ is the integral gain. The subscript $m$ represents the measured value. The terms $\omega L i_{q+}^+$, $-\omega L i_{d+}^+$, $-\omega L i_{q-}^-$, $+\omega L i_{d-}^-$ are inserted to decouple $dq$ axes dynamics.

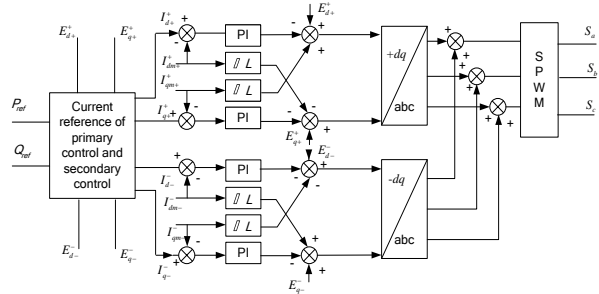

**Fig. 3.** Control structure of the dual current controller

As illustrated in Fig. 3, the dual current controller adjusts positive- and negative-sequence electrical quantity separately. The current commands appear as DC in rotate frame, and there is no need to build a tracking controller for AC signal. PWM signal is generated to control converter.

### B. Sampling of Positive- and Negative -Sequence Measurement

Positive- and negative-sequence currents should be separated from original sampling current, and dual current controller can work precisely. The monitored current is denoted in the stationary frame as follows:

$$I_{\alpha\beta m} = e^{j\omega t}I_{dqm+}^{+} + e^{-j\omega t}I_{dqm-}^{-} \tag{30}$$

By multiplying $I_{\alpha\beta m}$ by $e^{-j\omega t}$, (30) can be rewritten in positive synchronous reference frame as follows:

$$I_{dqm} = I_{dqm+}^{+} + e^{-j2\omega t}I_{dqm-}^{-} \tag{31}$$

In (31), the positive-sequence current is DC signal and negative-sequence current appears as double working frequency AC in positive reference frame. In the same way, the negative-sequence appears as DC in the negative reference frame, whereas the positive-sequence appears as double working frequency AC. Notch filters are adopted to separate positive- and negative-sequence currents and they can remove a narrow band of frequencies from the signal path. The transfer function of notch filters can be described as

$$G(s) = \frac{s^2 + \omega_1^2}{s^2 + 2\xi\omega_1 s + \omega_1^2} \tag{32}$$

where $\omega_1 = 200\pi$, the damping ratio $\xi = 0.25$. With the increase of digital control, the digital notch filters adopting sampled data theory should be design to meet the requirement. The bilinear transform is widely used for translating continuous domain transfer function into z-domain transfer function. The desired transfer function of notch filters in z-domain can be expressed as follows:

$$G(z) = \frac{1 - (2\cos\omega T)z^{-1} + z^{-2}}{1 + \xi\sin\omega T - (2\cos\omega T)z^{-1} + (1 - \xi\sin\omega T)z^{-2}} \tag{33}$$

where the sampling frequency $T = 10\text{kHz}$. According to transfer function, a recursive difference equation can be expressed as follows:

$$E_0(q) = [C_1 E_i(q) + C_2 E_i(q-1) + C_3 E_i(q-2) - C_5 E_0(q-1) - C_6 E_0(q-2)]/C_4 \tag{34}$$

where $E_i$ is the input, $E_0$ is the output, and $q$ is sampling numbers. Coefficients in (34) can be expressed as: $C_1 = C_3 = 1$, $C_2 = C_5 = -2\cos\omega T$, $C_4 = 1 + \xi\sin\omega T$, $C_6 = 1 - \xi\sin\omega T$.

The frequency and phase responses of notch filters are shown in Fig. 4. Obviously, the theoretical attenuation at the notch frequency is large, round off within discretization process weakens the actual realization of this amount of attenuation, but filtering capability of digital notch filters is maintained.

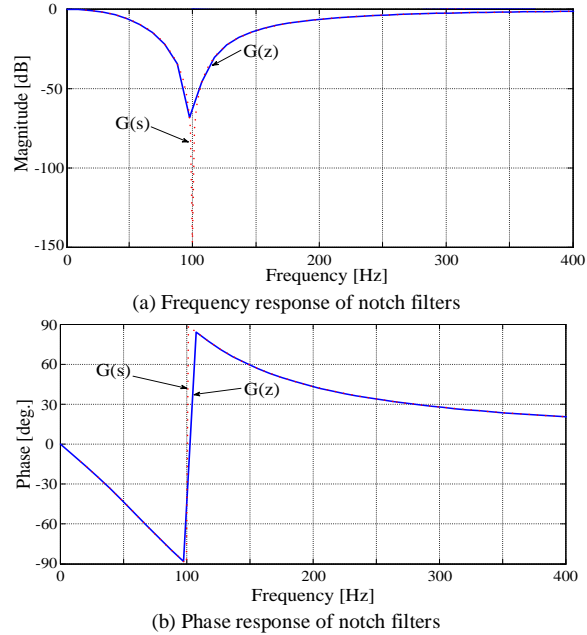

(a) Frequency response of notch filters

(b) Phase response of notch filters

**Fig. 4.** Bode diagram of the notch filters

The amplitude of negative-sequence electric quantity is much less than the magnitude of the positive-sequence electric quantity. When the positive-sequence current $I_{dqm+}^{+}$ is measured by notch filters, the negative-sequence current $I_{dqm-}^{-}$ can be separated as illustrated in Fig. 5.

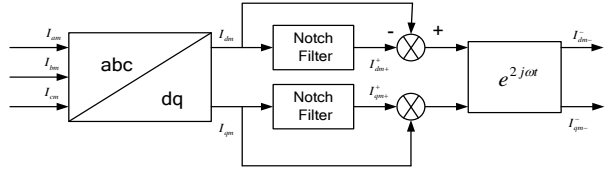

**Fig. 5.** Separation method for unbalanced current measurement

*C. Improved phase locked loop (PLL) technology*

In order to ensure the accuracy of phase measurement, phase locked loop is widely adopted in tracking phase. 100 Hz AC under synchronous reference frame plays negative effects when three-phase voltage is in an unbalanced state. If measurement of 100Hz AC voltage delays 1/4 frequency cycle and combines with the original value, negative effects of 100Hz AC can be eliminated. As illustrated in Fig. 6, no filters are adopted in this structure of PLL and phase lags which produces by inertia of filters does not exist.

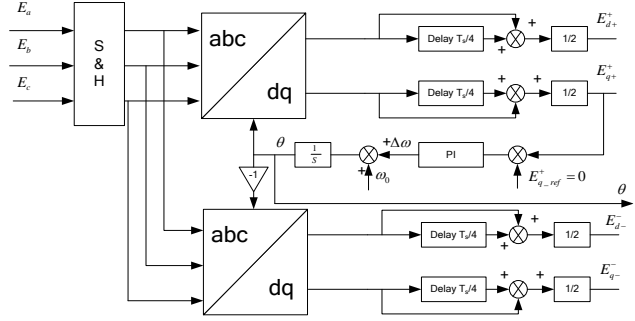

**Fig. 6.** Structure diagram of improved PLL

## VI. CASE STUDIES

In order to verify the performance of the proposed scheme, simulation studies are carried out in MATLAB/SIMULINK environment with a simulation time step of $2\ \mu s$. Parameters of the DG are given in Table 1, the power circuit of the converter and control structure are depicted in Fig. 1 and Fig. 3, respectively. Unbalanced three-phase input voltage at PCC $[E_a, E_b, E_c] = [341\sin(\omega t + 90°), 291\sin(\omega t - 30°), 311\sin(\omega t + 210°)]$ is illustrated in Fig. 7.

**Table 1** Parameters of the DG

| Parameters | Value |
| --- | --- |
| Fundamental frequency | $f_s = 50\text{Hz}$ |
| Switching frequency | $f_{sw} = 10\text{kHz}$ |
| DC bus voltage | $U_{dc} = 800\text{V}$ |
| Capacitance of DC bus | $C_1 = 8800\mu\text{F}$ |
| Filter inductance | $L = 5\text{mH}$ |
| Rated power | $S = 12\text{kVA}$ |
| Load parameter | $Z_1 = 10\Omega$ |

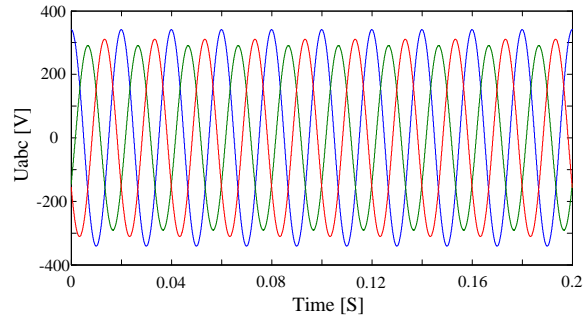

**Fig. 7.** Unbalance voltage at PCC

The process for stabilizing the active and reactive power oscillations is illustrated in Fig. 8. Active and reactive power references of the DG are $P_{ref} = 8\text{kW}$ and $Q_{ref} = 6\text{kVar}$. At the beginning of the simulation, primary control of the DG starts to eliminate active power oscillations. As illustrated in Fig. 8(a), the active power oscillations are effectively suppressed by primary control, but the amplitude of reactive power oscillations is up to 780Var. The large amplitude oscillations of reactive power can be explained by the special reciprocal relationship between active and reactive power oscillations in (12).

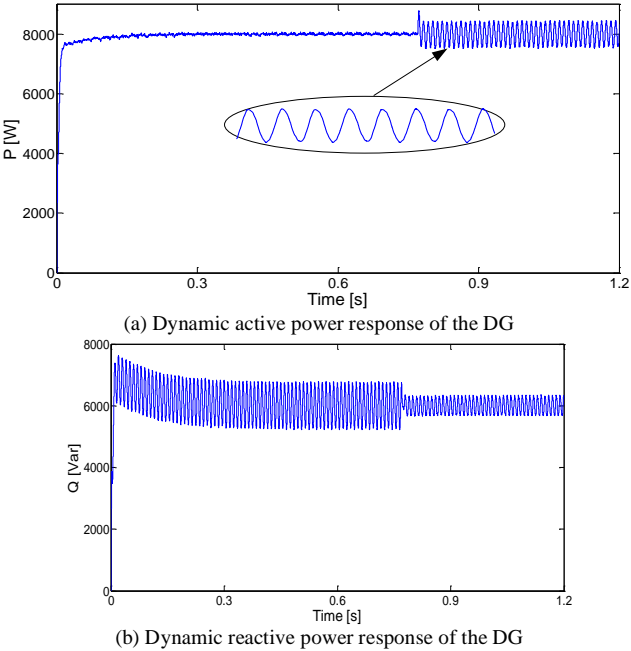

**Fig. 8.** Dynamic power response from primary control to secondary control

Secondary control is not activated until $t=0.77s$. As shown in Fig. 8(b), the amplitudes of active and reactive power oscillations are less than 450VA though the active power oscillations are amplified. Furthermore, the amplitudes of active and reactive power oscillations are proportional to respective power references and the power oscillation rate is limited to 5.6%. In addition, the injection of the negative-sequence current is significantly reduced under secondary control as illustrated in Fig. 9(a).

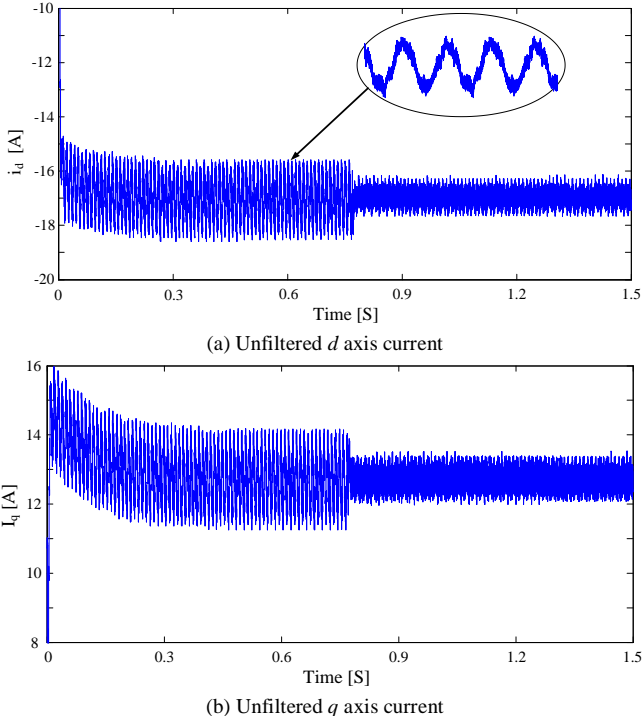

(a) Unfiltered *d* axis current

(b) Unfiltered *q* axis current

**Fig. 9.** Unfiltered *dq* current

Fig. 9 also shows unfiltered *dq* current of the DG. Amplitude of 100Hz AC component is large when secondary control is not activated. After secondary control starts, AC component is greatly weakened. After separation by notch filters, the

positive- and negative-sequence *dq* currents in Fig. 10 reflect characteristics of power oscillation amplitudes. The positive-sequence current is mainly proportional to the average output active power and reactive power, so it changes slightly when the control mode switches. The variation of negative-sequence current mainly determines the change of power oscillation amplitudes.

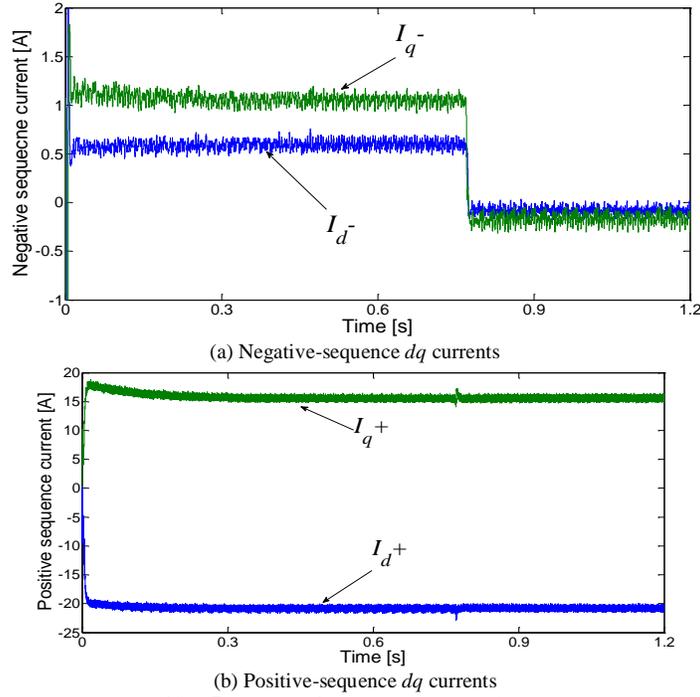

(a) Negative-sequence *dq* currents

(b) Positive-sequence *dq* currents

**Fig. 10.** Positive- and negative-sequence *dq* currents

The simulations results show that primary control suppresses active power oscillations at the cost of reactive power oscillations and negative-sequence current. In contrast, secondary control simultaneously reduces active and reactive power oscillations.

The voltage unbalance factor $\varepsilon_u$ [20] is defined as

$$\varepsilon_u = \frac{U_-}{U_+} \tag{35}$$

where $U_+$ and $U_-$ are respectively the positive- and negative-sequence voltages, and they can be obtained by symmetrical components calculation.

In the previous case, $\varepsilon_u$ is set to 4.6%. Increase the voltage unbalance factor to 9.5% at PCC and repeat the simulations. Active and reactive power references of the DG are $P_{ref} = 5\text{kW}$ and $Q_{ref} = 3\text{kVar}$. Due to the increase of voltage unbalance factor, reactive power oscillation rate is close to 31.7% under primary control. Once secondary control is activated, active and reactive power oscillation rates are limited to a reasonable level at the cost of a slight increase in active power oscillation rate.

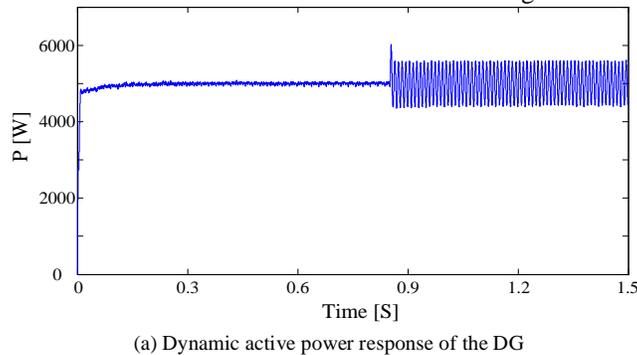

(a) Dynamic active power response of the DG

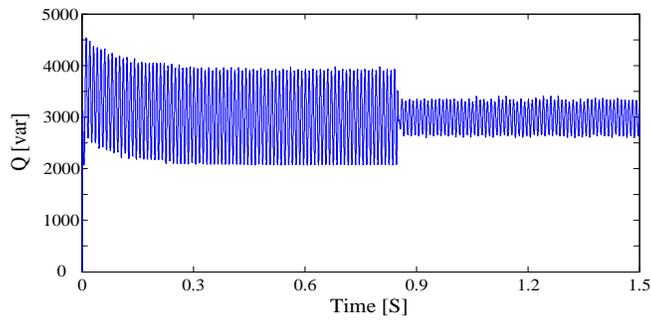

(b) Dynamic reactive power response of the DG
**Fig. 11.** Dynamic power response under high voltage unbalance factor

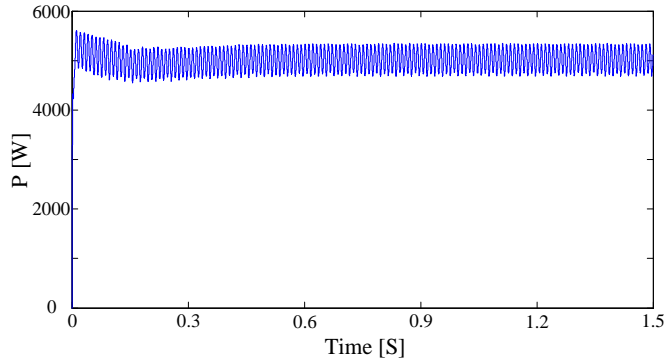

(a) Dynamic active power response of the DG

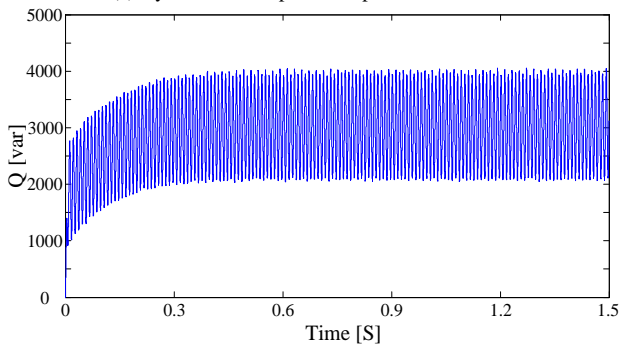

(b) Dynamic reactive power response of the DG
**Fig. 12.** Dynamic power response under the scheme introduced in [21]

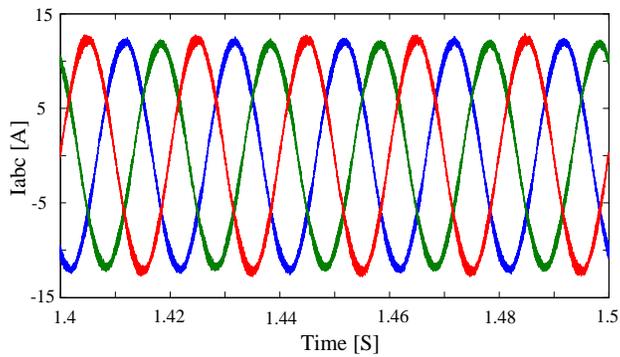

(a) Dynamic current response of the DG under secondary control

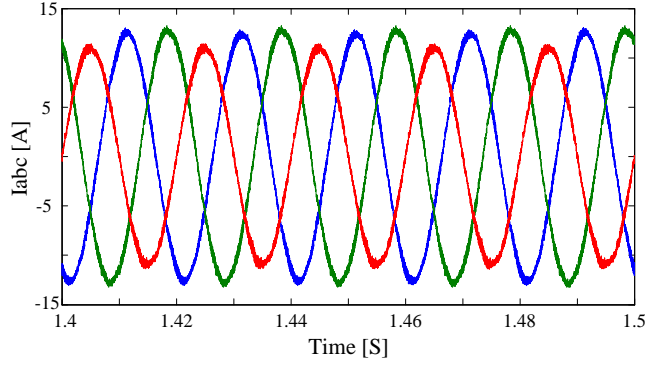
(b) Dynamic current response of the DG under the scheme introduced in [21]
**Fig. 13.** Simulation waves of three–phase current response

Fig. 12 shows power responses of negative-sequence disturbance compensation method introduced in [21]. The active power oscillations caused by this method is slightly less than the scheme proposed in this paper, but the reactive power oscillations are much severer. Therefore, the overall level of power oscillations in scheme introduced in [21] are greater than that in the proposed scheme. As shown in Fig. 13, the output current of proposed secondary control is better than scheme introduced in [21]. Furthermore, the current unbalance factor ($\varepsilon_I = I^- / I^+$) of secondary control is approximately 2.6% compared with 12.40% of the scheme in [21].

From the above results, secondary control causes active power oscillations, and active power oscillations lead to voltage ripple in DC link bus of the converter. The relationship between the DC voltage ripple and the active power oscillations can be expressed as [22]:

$$V_{\text{ripple}} = \frac{P_v}{2C_1 U_{dc} \omega} \tag{36}$$

where $C_1$ is equivalent capacitance of DC-link bus. The amplitudes of active power oscillations which are caused by secondary control in above two cases are 446W and 600W, respectively. By using (36), DC voltage ripple is about 0.1V and 0.136V respectively. Therefore, the voltage ripple has slight influence on the DC-link bus.

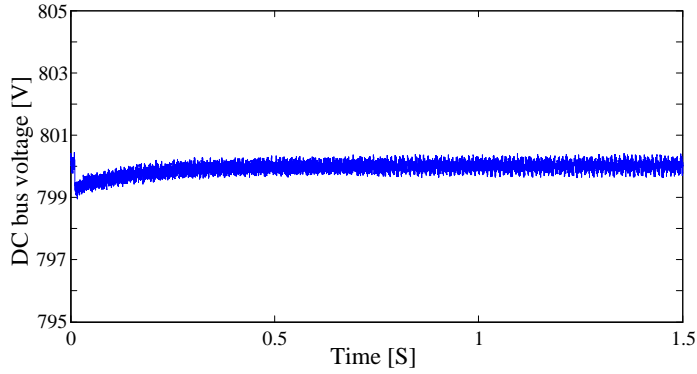
(a) DC bus voltage under low voltage unbalance factor

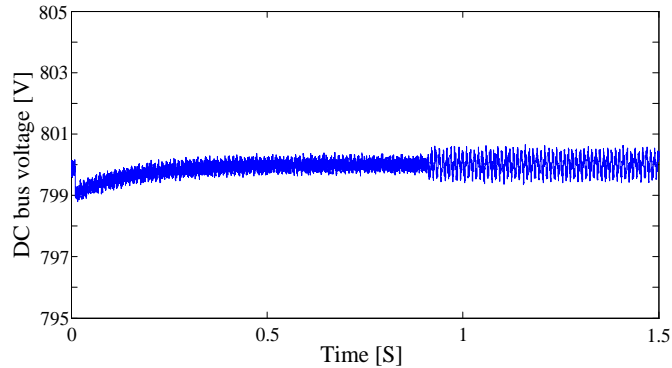
(b) DC bus voltage under 23% voltage unbalance factor
**Fig. 14.** DC bus voltage of the DG converter

Fig. 14 shows the DC bus voltage response under voltage unbalance factor 4.6% and 23%. The DC voltage undergoes the process from a transient dip to recovery when the DG starts. The amplitude of voltage ripple can be ignored under secondary control in case of low voltage unbalance factor. Furthermore, the amplitude of voltage ripple is relatively smaller under 23% voltage unbalance factor.

## VII. Conclusion

This paper focuses on power oscillations of DGs under unbalanced voltage. The analytic relationship between active and reactive power oscillations is revealed, and a hierarchical control scheme is proposed to control power oscillations of DGs. Based on the optimal model, the amplitudes of the power oscillations can be precisely controlled by the injected currents whose references are calculated by using QNTR method. Simulation results show that both active and reactive power oscillations of the DG can be accurately limited and the injected negative-sequence currents are reduced effectively. Therefore, the proposal befits power oscillations reduction for DGs under unbalanced working conditions.

## Acknowledgements

This work was supported in part by the National Natural Science Foundation of China (51377016), the Science and Technology Project of State Grid Corporation of China under Grant No. 2015GW-90 (research on interaction mechanism and coordinated control technology of flexible HVDC and AC electric grid) and the Doctor Scientific Research Foundation of Northeast Dianli University (BSJXM-201407).